\documentstyle[12pt]{report}
\title{Derivation of Non-isotropic Phase Equations from
a General Reaction-Diffusion Equation}
\author{Y.Masutomi and K. Nozaki\\
        Department of Physics, Nagoya University\\
          Nagoya 464-8602, Japan}

\makeindex
\begin{document}

\baselineskip 2.5ex
\parskip 3.5ex
\maketitle

\renewcommand{\thesection}{\arabic{section}}
\renewcommand{\thesubsection}{\arabic{section}.\Alph{subsection}}
\renewcommand{\theequation}{\arabic{section}.\arabic{equation}}
\newcommand{\eps}{\epsilon}
\newcommand{\A}{\tilde A}
\newcommand{\B}{\tilde B}
\newcommand{\p}{\tilde p}
\newcommand{\h}{\tilde h}
\newcommand{\tP}{\tilde P}
\newcommand{\tU}{\tilde U}
\newcommand{\pvt}{\tilde{\bm p}}
\newcommand{\pv}{\bm p}
\newcommand{\rv}{\bm r}
\newcommand{\etav}{\bm {\eta}}
\newcommand{\phit}{\tilde{\phi}}
\newcommand{\rvp}{{\bm r}_{\perp}}
\newcommand{\ad}{\dot a}
\newcommand{\add}{\ddot a}
\newcommand{\bm}[1]{\mbox{\boldmath $#1$}}
\newcommand{\svec}[1]{\mbox{{\footnotesize $\bm{#1}$}}}
\renewcommand{\vec}{\bm}
\newcommand{\der}[2]{\frac{d {#1}}{d {#2}}}
\newcommand{\beq}{\begin{equation}}
\newcommand{\beqa}{\begin{eqnarray}}
\newcommand{\eeq}{\end{equation}}
\newcommand{\eeqa}{\end{eqnarray}}
\newcommand{\la}{\langle}
\newcommand{\ra}{\rangle}
\renewcommand{\vec}{\bm}
\newcommand{\un}{\underline}
\newcommand{\noi}{\noindent}
\newcommand{\lb}{\label}
\newcommand{\fr}[1]{(\ref{#1})}
\newcommand{\non}{\nonumber}
\newcommand{\fns}{\footnotesize}
\newcommand{\map}{\rightarrow}
\newcommand{\maps}{\rightarrow}
\newcommand{\imply}{\Rightarrow}
\newcommand{\implies}{\Rightarrow}
 
\begin{abstract}

A non-isotropic version of phase equations such as the Burgers equation,
the  K-dV-Burgers equation, the Kuramoto-Sivashinsky equation and the
Benney equation in the three-dimensional space is
 systematically derived from  a general reaction-diffusion system
 by means of the renormalization group method.


\end{abstract}

\section{Introduction}

It was a long time ago  that the Kuramoto-Sivashinsky (K-S)
 equation was proposed as a higher-order
phase equation describing an unstable phase state \cite{kura}.
Although there have been many works based on the K-S equation since then,
 there are only  few derivations of the K-S equation based
 on the singular perturbation methods.
 Recently, the isotropic K-S equation has been derived as a phase
 equation of a periodically oscillating solution of the complex
 Ginzburg-Landau equation by means of the renormalization
 group method \cite{maru}.
 In a reaction-diffusion system,
   spatial symmetry  often breaks so that both spatially and
 temporally oscillating solutions emerge. In such a case, it is anticipated
 that  a slowly varing
 phase of  the symmetry-breaking oscillating state is asymptotically
 governed by a non-isotropic phase equation.
 The Gross-Newell's phase equation was derived by the RG method \cite{sasa}.
 However, there are no
 explicit derivations of such non-isotropic phase equations
 as the non-isotropic K-S equation, the K-dV-Burgers equation and
 the Benney equation \cite{benny} from a general reaction-diffusion equation.
 In this paper, we derive such non-isotropic phase equations to a
 symmetry-breaking state of a general reaction-diffusion equation by
 means of the renormalization group (RG) method.\\
\section{Renormalization Group Method}
The perturbative RG method
introduced in  \cite{cgo} is shown to be
interpreted as the procedure
to obtain an asymptotic expression of a generator of a renormalization
transformation based on the Lie group \cite{gmn}.
This Lie group approach provides the following simple recipe for obtaining
an asymptotic form of a RG equation from  ordinary differential equations
(ODE).\\
(1) Get a secular series solution of a perturbed equation by means of naive
perturbation calculations.\\
(2)Find integral constants, which are renormalized to elliminate all the
secular terms in the perturbed solution and give a renormalization
transformation for the integral constants.\\
(3)Rewrite the renormalization transformation by excuting an arbitrary
shift operation on the independent variable:$t\to t+\tau$ and derive
a representation of a Lie group underlying the renormalization
transformation.\\
(4)By differentiating the representation of the Lie group with respect to
arbitrary $\tau$, we obtain an asymptotic expression of the generator,
which yields an asymptotic RG equation.\\
This procedure is valid for general ODE regardless of  translational
symmetry.
The above recipe for ODE is also applicable to autonomous partial
differntial equations (PDE)
by choosing suitable polynomial kernels of the linearized operator
appearing in perturbed equations.
First, we should take the lowest-order polynomial, of which degree is one,
 as the leading order secular term. As perturbation calculations proceed to
 the higher order, polynomial kernels of higher degrees are included in
 the higher-order secular terms order by order. When we can continue
 this process consistently, we say that secular terms of polynomials are
 renormalizable or ,simply, the consistent renormalization condition is
 satisfied in the sense of the Lie approach.
If this is the case,
we can determine suitable polynomial kernels among infinite
number of  kernels of the linearized operator and the step (1)
in the recipe is completed.  There are no problems in the other steps.
Thus, the Lie-group approach is consistently applicable to PDE
and derivation of some soliton equations and simple phase
equations (e.g. the isotropic Burgers equation)  was  presented in
\cite{gmn}.\\
In the following sections, using this Lie-group approach of the RG method,
we derive various non-isotropic phase equations from a general reaction
diffusion system.
\section{Non-isotropic Burgers Equation}
\quad Let us consider a general
reaction-diffusion system of equations:
\beq
\partial_t U=F(U)+D\nabla^2U,\lb{nb1}
\eeq
where $U$ is an $n$-dimensional vector and $D$ is an $n\times n$ constant
matrix. Suppose \fr{nb1} has a spatially and temporally oscillating
solution  $U_0=U_0(k,kx-\omega(k)t+\phi)$ satisfying
\beq
-(\omega U_{0,\theta}+F(U_0)+k^2DU_{0,\theta \theta})=0, \lb{nb2}
\eeq
where $k$ and $\phi$ are arbitrary constants;
 $\omega(k)$ is a definite function of $k$; $\theta=kx-\omega t+\phi$;
 the suffix $\theta$ denotes the derivative with respect to $\theta$ and
 $U_0$ is a $2\pi$ periodic function of $\theta$. For later convenience,
we list some useful identities. Differentiating \fr{nb2} with
respect to $\theta$ and $k$, we have
\beqa
L(\theta)U_{0,\theta}&\equiv&-(\omega \partial_\theta+F'(U_0)\cdot
+k^2D\partial^2_\theta)U_{0,\theta}=0,
\lb{nb3}\\
LU_{0,\theta \theta}&=&F'': U^2_{0,\theta},\lb{nb4}\\
LU_{0,k}&=&( \dot{\omega}+2kD\partial_{\theta})U_{0,\theta}
\equiv MU_{0,\theta},\lb{nb5}\\
LU_{0,kk}&=&F'': (U_{0,k})^2+2MU_{0,\theta k}
+2DU_{0,\theta \theta}+\ddot{\omega}U_{0,\theta},\lb{nb6}\\
LU_{0,\theta k}&=&MU_{0,\theta \theta}+F'': U_{0,\theta}U_{0,k},\lb{nb7}
\eeqa
where $F'\cdot V=(V\cdot \nabla_{U})F(U)|_{U=U_0}$,\quad
 $F'': VW=[(W\cdot \nabla_{U})(V\cdot \nabla_{U})F(U)]|_{U=U_0}$,
 $\dot{\omega}=\partial_k\omega$ and
$\ddot{\omega}=\partial^2_k\omega$.\\
 Let us seek a secular solution close to $U_0(k,\theta)$:
\beq
U=U_0(k+\kappa(x,\rvp,t),\theta +\delta(x,\rvp,t))
   +\tilde U(\theta,x,\rvp,t),\lb{solution}
\eeq
where $\rvp=(0,y,z)$; $\delta(x,\rvp,t)$ and $\kappa(x,\rvp,t)$ are
small secular deviations from the constant phase $\phi$ and
the wavenumber $k$
respectively and so
\beq
\partial_x\delta\equiv \delta_x=\kappa.\lb{kappa}
\eeq
 $\tilde U(\theta,x,\rvp,t)$ represents small perturbed fields which
 modifies the 0-th order field pattern $U_0$ and
 is not expressed by differentials of $U_0(k,\theta)$.
The arguments $(x,\rvp,t)$ designate ``secular variables'', that is,
all secular perturbed fields are polynomials with respect to  $(x,\rvp,t)$
and periodic with $\theta$.
An expansion of $U_0$  in terms of small deviations
$\delta$ and $\kappa$ yields
\beqa
U_0(k+\kappa,\theta +\delta)&=&U_0(k,\theta)+\delta U_{0,\theta}
+\kappa U_{0,k} \non\\
&+&(\delta^2/2)U_{0,\theta\theta}+(\kappa^2/2)U_{0,kk}
+\delta \kappa U_{0,\theta k}+\cdots.\non
\eeqa
Note that $\delta(x,\rvp,t)$ (and $\kappa$) is a secular or
 polynomial function of $(x,\rvp,t)$, which should be
 eliminated by renormalizing the phase $\phi$ later. \\
In this section we suppose that $\delta$ and $\tilde U$ are expanded in
terms of a small perturbation parameter $\eps$ as
\beqa
\delta&=&\eps(P_1+\eps P_2+\cdots),\quad
\kappa=\eps(P_{1,x}+\eps P_{2,x}+\cdots) \lb{delex1} \\
\tilde U&=&\eps^2(\tilde U_2(\theta)+\eps \tilde U_3+\cdots),\lb{uex1}
\eeqa
where the suffix $x$ denotes the derivative with respect to $x$ and
$P_j\quad (j=1,2,\cdots)$ are polynomials of  $(x,\rvp,t)$,
which have increasing degrees with $j$ so that polynomial secular
terms are renomalizable in the sense of the Lie approach of the RG method
\cite{gmn}.
$\tilde U_2$ depends only on $\theta$ since the leading order term of
 $\tilde U$ is not contain secular terms of $(x,\rvp,t)$
 ; otherwise they would not be eliminated by the RG procedure
  (see \fr{kbrex} and \fr{kbrrn} or \fr{kbr} and \fr{rren} ).
 Then \fr{solution} reads
 \beqa
 U&=&U_0(k,\theta)+\eps U_1+\eps^2 U_2+\cdots, \lb{nbexu}\\
 U_1&=&P_1U_{0,\theta}(k,\theta)+P_{1,x}U_{0,k}(k,\theta),\lb{nb9} \\
 U_2&=&P_2U_{0,\theta}+P_{2,x}U_{0,k}+(1/2)P^2_1U_{0,\theta \theta}\non\\
&&+P_1P_{1,x}U_{0,\theta k}+(1/2)P^2_{1,x}U_{0,k k}+\tilde U_2.\lb{nb12}
 \eeqa
 Introducing the Galilean transformation
\beq
\quad x'=x-\dot \omega t,\quad t'=t, \non
\eeq
 and substituting \fr{nbexu} into \fr{nb1},
 we have to the first order perturbed field $U_1$
\beq
(\partial_{t'}+L(\theta)-\partial_{x'}M-\nabla^2D)U_1\equiv \tilde{L}U_1=0,
\lb{nb8}
\eeq
where $( \theta,x',t',\rvp)$ is considered as a set of independent
variables.
Hereafter, the prime attached to $(x',t')$ is omitted for simplicity.
Substituting \fr{nb9} into \fr{nb8} and using \fr{nb3} and \fr{nb5}, we have
\beqa
&&-P_{1,xx}MU_{0,k}
-\nabla^2P_1DU_{0,\theta}
-\nabla^2P_{1,x}DU_{0,k}\non\\
&&+P_{1,t}U_{0,\theta}
+P_{1,xt}U_{0,k}=0.\lb{nb10}
\eeqa
Since $P_1$ is a polynomial of $(x,\rvp,t)$ and $U_0$ is a periodic
function of $\theta$ , \fr{nb10} reads
\beq
P_{1,xx}=0,\quad \nabla^2P_1=0,\quad \nabla^2P_{1,x}=0,\quad
 P_{1,t}=0,\quad P_{1,xt}=0,\lb{burp}
\eeq
where the suffix $t$ denotes the derivative with respect to $t$. Noting
that the leading order secular term consists of a polynomial of degree one,
\fr{burp} yields
\beq
P_1=P_{1,x}x+\nabla_{\perp}P_1\cdot \rvp.\lb{bg1}
\eeq
where $\nabla_{\perp}\equiv
(0,\partial_y,\partial_z)$; $P_{1,x}$ and $\nabla_{\perp}P_1$ are
arbitrary constants.\\
 The second order equation obeys
\beq
\tilde{L}U_2=(1/2)F'': U_1^2(\theta), \lb{nb11}
\eeq
Substituting \fr{nb12} into \fr{nb11},
 we obtain with the aid of \fr{nb3}--\fr{nb7}
\beqa
 L \tilde U_2&=&
|\nabla_{\perp}P_1|^2DU_{0,\theta \theta}(\theta)
-(1/2)P^2_{1,x}\ddot{\omega}U_{0,\theta}(\theta)+P_{2,xx}MU_{0,k}(\theta)\non\\
&&+\nabla^2P_2DU_{0,\theta}(\theta)-P_{2,t}U_{0,\theta}(\theta)
-P_{2,x,t}U_{0,k}(\theta),\lb{nib8}
\eeqa
which is  an equation for periodic $\tilde U_2$.
Since $\tilde U_2$ is a
function of $\theta$ only, all the coefficients of functions of $\theta$
in the right hand side (RHS) of \fr{nib8} do not depend on
$(x,\rvp,t)$, that is,
\beq
P_{2,xx}=c_1,\quad \nabla^2 P_2=c_2,\quad P_{2,t}=c_3,\quad P_{2,xt}=c_4,
\non
\eeq
where $c_1,c_2$ and $c_3$ are non-zero constants while $c_4=0$ due to the
consistent renormalization condition, i.e. both $t$ and $xt $
do not enter in $P_2$ as secular terms to be removed consistently by
renormalization. This requirement holds throught this paper.
An explict form of $P_2$ is given as a polynomial
of degree two with respect to $(x,\rvp)$:
\beq
P_2=P_{2,xx}x^2/2+(\rvp \rvp: \nabla_{\perp} \nabla_{\perp})P_2/2
+x(\rvp:\nabla_{\perp})P_{2,x}+P_{2,t}t, \lb{bg2}
\eeq
where all the coefficients of monomials, i.e., $P_{2,xx},
\nabla_{\perp}\nabla_{\perp} P_2,\nabla_{\perp} P_{2,x}$ and $P_{2,t}$ are
arbitrary constants and
\beq
(\overbrace{\rvp \cdots \rvp}^n :\overbrace{\nabla_{\perp} \cdots
 \nabla_{\perp}}^n)\equiv \sum_{k=0}^{k=n}
\left( \begin{array}{c} n\\ k \end{array} \right)
y^{n-k} z^k \partial^{n-k}_y \partial^{k}_z.
\eeq
Then, a periodic solution $\tilde U_2$ is possible only if the following
compatibility condition is satisfied.
\beqa
&&|\nabla_{\perp}P_1|^2\la \hat{U}\cdot DU_{0,\theta \theta}\ra
-(1/2)\ddot{\omega}(P_{1,x})^2\la \hat{U}\cdot U_{0,\theta}\ra
+P_{2,x x}\la \hat{U}\cdot MU_{0,k}\ra \non\\
&&+\nabla^2P_2\la \hat{U}\cdot DU_{0,\theta}\ra
-P_{2,t}\la \hat{U}\cdot U_{0,\theta}\ra =0,\lb{comnb}
\eeqa
where $\hat{U}$ is an adjoint function of a null eigenfunction of $L$
and $<\hat{U}\cdot U>\equiv\int_{0}^{2\pi}(\hat{U}\cdot U)d\theta$.\\
\quad A secular solution up to $O(\eps^2)$ is
\beqa
U=U_0(k,kx-\omega t+\phi)&+&\eps( P_1+\eps P_2)U_{0,\theta}
+(1/2)\eps^2P^2_1U_{0,\theta \theta}\non\\
&+&\eps( P_{1,x}+\eps P_{2,x})U_{0,k}+(1/2)\eps^2(P_{1,x})^2U_{0,k k}\non\\
&+&\eps^2P_1P_{1,x}U_{0,\theta k}\non\\
=U_0(k+\phit_x,kx-\omega t+\phit),
\eeqa
where $\phit$ is a renormalized phase defined by
 a renormalization transformation
\beqa
\phit(x,\rvp,t)=\phi+\delta(x,\rvp,t)
=\phi+\eps(P_1+\eps P_2+\cdots).\lb{nb14}
\eeqa
 Since the renormalized phase should enjoy translational symmery with
 respect to  independent variables, \fr{nb14} is rewritten as
, shifting the origin $(x,\rvp,t)=(0,0,0)$ to an arbitrary point
 $(x,\rvp,t)$,
\beqa
\phit(t+\tau,x+\xi,\rvp+\etav&)&=\phi(x,\rvp,t)\non\\
&+&\eps\{\tilde P_1(\xi,\etav;x,\rvp,t)+
\eps\tilde P_2(\xi,\etav,\tau;x,\rvp,t)\},\lb{nb15}
\eeqa
where $(x,\rvp,t)$ in polynomials $P_1$ and $P_2$ are replaced by
$(\xi,\etav,\tau)$
and their coefficients depend on the coordinate of the origin $(x,\rvp,t)$
,e.g.
$$\tilde P_1(\xi,\etav;x,\rvp,t)=P_{1,x}(x,\rvp,t)\xi+
\nabla_{\perp} P_1(x,\rvp,t)\cdot\etav$$ and so on.
This reinterpretation of  coefficients of secular terms is
the key ingredient of the Lie approach of the RG method \cite{gmn}.
Hence, Eq.\fr{nb15} with \fr{bg1} and \fr{bg2} reads
\beqa
\phit_{x}&=&[\partial_{\xi}(\eps\tP_1+\eps^2\tP_2)]_0=\eps P_{1,x},\non\\
\nabla_{\perp}\phit&=&[\partial_{\etav}(\eps\tP_1+\eps^2\tP_2)]_0
=\eps\nabla_{\perp} P_1, \non\\
\phit_{x x}&=&[\partial^2_{\xi}(\eps^2\tP_2)]_0=\eps^2 P_{2,xx},\non\\
\nabla^2_{\perp}\phit&=&[\partial^2_{\etav}(\eps^2\tP_2)]_0=
\eps^2\nabla^2_{\perp}P_2,\non\\
\phit_{t}&=&[\partial_{\tau}(\eps^2\tP_2)]_0=\eps^2P_{2,t},\lb{rgnb}
\eeqa
where $[f(\xi,\etav,\tau;x,\rvp,t)]_0=f(0,0,0;x,\rvp,t)$.
Substituing these relations between differentials
of the renormalized phase $\phit$ and the reinterpreted coefficients of
polynomials $P_1$ and $P_2$  into \fr{comnb},
 we obtain a non-isotropic Burgers (n-Burgers) equation:
\beqa
\phit_{t}=D_{\parallel}\phit_{x x}+D_{\perp}\nabla^2_{\perp}\phit
+N_{\parallel}(\phit_{x})^2+N_{\perp}|\nabla_{\perp}\phit|^2,\lb{nib}
\eeqa
and
\beqa
D_{\parallel}&=&D_{\perp}+D'_{\parallel},\non\\
D_{\perp}&=&\la \hat{U}\cdot DU_{0,\theta}\ra
/\la \hat{U}\cdot U_{0,\theta}\ra,
\quad D'_{\parallel}=\la \hat{U}\cdot MU_{0,k}\ra
/\la \hat{U}\cdot U_{0,\theta}\ra \non\\
N_{\perp}&=&\la \hat{U}\cdot DU_{0,\theta \theta}\ra
/\la \hat{U}\cdot U_{0,\theta}\ra \non\\
N_{\parallel}&=&N_{\perp}
+\{(1/2)\la \hat{U} \cdot F'': (U_{0,k})^2\ra +\la \hat{U}
\cdot MU_{0,\theta k}\ra \}/ \la \hat{U}
 \cdot U_{0,\theta} \ra,\non\\
 &=&-\ddot{\omega}/2 ,\lb{nib15}
\eeqa
where  the last equality of \fr{nib15} comes from \fr{nb6}.

\section{K-dV-Burgers Equation}

\quad In this section, it is assumed that the diffusion coefficient
$D_{\parallel}$ along the $x$ direction in the n-Burgers
equation \fr{nib} is small as $\epsilon$:
\beq
D_{\parallel}\propto \la \hat{U}\cdot (DU_{0,\theta}+ MU_{0,k})\ra
\sim O(\eps).\lb{kb1}
\eeq
Nevertheless the net diffusion in the $x$ direction is supposed to be
much greater than that in the $\rvp$ direction, i.e.
\beq
\nabla_{\perp}/\partial_x \sim O(\eps).\lb{scalekdv}
\eeq
 Eq.\fr{kb1} implies that there is a periodic vector $V(\theta)$
such that
\beq
LV(\theta)=DU_{0,\theta}+MU_{0,k}+ O(\eps).\lb{kb2}
\eeq
 Suppose that $\delta , \tilde U$, and $U$ are expanded as
\beqa
\delta&=&\eps^2\{P_1(x,\rvp,t)+\eps P_2(x,\rvp,t)+\cdots\},\lb{kb}\\
\tilde U&=&\eps^3\{\tilde U_2(\theta)+\eps \tilde U_3(\theta,x,\rvp,t)
+\cdots\},\lb{uex2}\\
U&=&U_0(k,\theta)+\eps^2( U_1+\eps U_2+\eps^2 U_3+\cdots), \lb{kbux}
\eeqa
 then the leading order perturbed terms
($O(\eps^2)$) gives the same equations as \fr{nb8}--\fr{burp}.
As a polynomial solution of \fr{burp}, we choose $P_1$ as
\beq
P_1=P_{1,x}x,\quad P_{1,x}=\mbox{constant},\lb{kbsec1}
\eeq
instead of \fr{bg1} since $|P_{1,x}|\gg|\nabla_{\perp}P_1|$ due to the
assumption \fr{scalekdv}.\\
 To $O(\eps^3)$, we have
\beqa
\tilde{L}U_2&=&0,\lb{kb31}\\
U_2&=&P_2(x,\rvp,t)U_{0,\theta}(\theta)+P_{2,x}(x,\rvp,t)U_{0,k}(\theta)
+\tilde U_2(\theta),
\lb{kb32}
\eeqa
Substituting \fr{kb32} into \fr{kb31}, we get
\beqa
\tilde{L}\tilde U_2&=&P_{2,xx}(MU_{0,k}+DU_{0,\theta})
+\nabla_{\perp}^2 P_{2}DU_{0,\theta}
+\nabla^2P_{2,x}DU_{0,k}\non\\
&&-P_{2,t}U_{0,\theta}-P_{2,xt}U_{0,k}.\lb{kb3}
\eeqa
Due to the consistent renormalization condition described in section 2
 (i.e., consistent increasing of degrees of polynomial secular terms $P_j$),
$P_2$ is a polynomial of degree two with respect to $x$ and
of degree one with respect to $\rvp$ at most so that we can set
$\nabla_{\perp}^2P_2=\nabla^2P_{2,x}=0$ and $P_{2,xt}=0$,
while
\beq
P_{2,xx}=\mbox{constant}=R_2. \lb{kbtu1}
\eeq
If $ P_{2,t}\ne 0$, the fourth term of
the RHS of \fr{kb3} is only a term which causes a secular behaviour of
$\tU_2(\theta)$ with respect to $\theta$ and so $P_{2,t}=0$.
Then by virtue of \fr{kb2} we have
\beq
\tU_2(\theta)=R_2V(\theta),\lb{kb4}
\eeq
and
\beq
P_2=R_2x^2/2+\nabla_{\perp}P_2\cdot \rvp,\quad \nabla_{\perp}P_2=
\mbox{constant}.\lb{kbsec2}
\eeq
To $O(\eps^4)$ we have
\beqa
\tilde{L}U_3&=&(1/2)F'':(U_1)^2,\lb{kb9}\\
U_3&=&P_3(x,\rvp,t)U_{0,\theta}+P_{3,x}(x,\rvp,t)U_{0,k}
+(1/2)(P_1)^2U_{0,\theta \theta}\non\\
&&+P_1P_{1,x}U_{0,\theta k}
+(1/2)(P_{1,x})^2U_{0,k k}\non\\
&&+\tilde U_3(\theta,x,\rvp,t),\lb{kb10}
\eeqa
from which we get
\beqa
\tilde L \tilde U_3&=&P_{3,xx}MU_{0,k}
+\nabla^2P_3DU_{0,\theta}
-(1/2)(P_{1,x})^2\ddot{\omega}U_{0,\theta}\non\\
&&
+\nabla^2P_{3,x}DU_{0,k}-P_{3,t}U_{0,\theta}-P_{3,xt}U_{0,k}.\lb{kb5}
\eeqa
In view of \fr{kbsec1} and \fr{kbsec2}, the consistent renormalization
condition requires that $P_3$ contains $x^3$ and the first term of
the RHS of \fr{kb10} is secular with respect to $x$.
Therefore we set
\beq
\tilde U_3(\theta,x,\rvp,t)=R_3(x)V(\theta)+\bar U_3(\theta),\lb{kbtu2}
\eeq
where
\beq
R_3=R_{3,x}x,\quad R_{3,x}=\mbox{constant},
\eeq
and  \fr{kb5} is rewritten as
\beqa
L \bar U_3(\theta)&=&-(R_3-P_{3,xx})(MU_{0,k}+DU_{0,\theta})
+\nabla_{\perp}^2P_3DU_{0,\theta}
-(1/2)(P_{1,x})^2\ddot{\omega}U_{0,\theta}\non\\
&&-R_{3,x}MV(\theta)
+\nabla^2P_{3,x}DU_{0,k}-P_{3,t}U_{0,\theta}-P_{3,xt}U_{0,k}.\lb{kb13}
\eeqa
Since \fr{kb13} is an equation to $\bar{U_3}(\theta)$,
all the coefficients of functions of $\theta$ in the RHS of \fr{kb13}
do not depend on $(x,\rvp,t)$,that is,
\beqa
&&R_3-P_{3,xx}=c_1,\quad \nabla_{\perp}^2P_3=c_2,\non\\
&& \nabla^2P_{3,x}=c_3,\quad P_{3,t}=c_4,\quad P_{3,xt}=c_5,\lb{kbsec3}
\eeqa
where $c_n$ are arbitrary constants. The consistent renormalization
condition yields $c_2=c_5=0, c_3=P_{3,xxx}$ and $c_1=0$ so that
a secular coefficient $R_3(x)$ is consistently removed by a renormalization
transformation to $\phi$ (see \fr{kbrrn}). Thus \fr{kbsec3} gives
\beq
P_3=P_{3,xxx}x^3/3!+x(\rvp:\nabla_{\perp})P_3+P_{3,t}t, \lb{kbp3}
\eeq
The compatibility condition for a periodic solution
$\bar U_3(\theta)$
requires that
\beqa
&&P_{2,x x}(\la \hat{U}\cdot MU_{0,k}\ra
+\la \hat{U}\cdot DU_{0,\theta}\ra)/\eps
 -(\ddot{\omega}/2)(P_{1,x})^2 \la\hat{U}\cdot U_{0,\theta}\ra \non\\
&&+P_{3,x x x}(\la \hat{U}\cdot MV\ra+
 \la \hat{U}\cdot DU_{0,k}\ra)
-P_{3,t}\la \hat{U}\cdot U_{0,\theta}\ra =0.\lb{comkb}
\eeqa
Here the first term in the LHS of \fr{comkb} comes from  $O(\eps^3)$
terms.\\
 Now we arrive at a renormalization transformation to $O(\eps^4)$
\beq
\phit(x,\rvp,t)=\phi
+\eps^2\{P_1(x)+\eps P_2(x,\rvp)+\eps^2P_3(x,\rvp,t)\}.\lb{kbrnm}
\eeq
Following the same procedure as in the case of the n-Burgers equation,
 we obtain from \fr{kbrnm},\fr{kbsec1},\fr{kbsec2} and \fr{kbp3}
\beq
\phit_{x}=\eps^2 P_{1,x},\quad
\phit_{x x}=\eps^3P_{2,xx},\quad
\phit_{x x x}=\eps^4P_{3,xxx},\quad
\phit_{t}=\eps^4P_{3,t}.\lb{rgkb}
\eeq
Substituting \fr{rgkb} into \fr{comkb}, we arrive at
 the K-dV-Burgers (K-B) equation:
\beq
\phit_{t}=D_{\parallel}\phit_{x x}
+N_{\parallel}(\phit_{x})^2+A\phit_{x x x},\lb{nkb}
\eeq
where
\beqa
A&=&B+A',\non\\
B&=&\la \hat{U}\cdot MV\ra /
\la \hat{U}\cdot U_{0,\theta}\ra,\quad
A'=\la \hat{U}\cdot DU_{0,k}\ra /
\la \hat{U}\cdot U_{0,\theta}\ra. \lb{cofkb}
\eeqa
Notice that secular terms in $\tU$ are also eliminated by the present
renormalization procedure.
To $O(\eps^4)$, \fr{kb4} and \fr{kbtu2} give
\beq
\tU=\eps^3\{R_2+\eps R_3(x))\}V(\theta)+
\eps^4 \bar U_3(\theta),\lb{kbrex}
\eeq
where  a secular term $R_3$ is removed by introducing
a renormalized $R_2$ such that
\beq
\tilde R_2(x,\rvp,t)=R_2+\eps R_3(x).\lb{kbrrn}
\eeq
This renormalization transformation is found to be consistent
with that to the phase $\phi$ \fr{kbrnm}, since \fr{kbtu1}, \fr{rgkb}
 and \fr{kbsec3} with $c_1=0$ imply
$\tilde R_2=\phit_{xx}/\eps^3$.

If we introduce the following expansion
\beqa
\delta&=&\eps^4\{P_1+\eps P_2+\cdots\},\non\\
\tilde U&=&\eps^5\{\tilde U_2+\eps \tilde U_3
+\cdots\},\non\\
U&=&U_0(k,\theta)+\eps^4( U_1+\eps U_2+\eps^2 U_3+\cdots), \non
\eeqa
instead of \fr{kb},\fr{uex2},\fr{kbux}, and an auxiliary condition $D_{\parallel}\sim O(\eps^2)$ instead of \fr{kb1},
the similar procedure as above yields , to $O(\eps^8)$, the following
K-dV-Burgers equation with the perpendicular diffusion term
\beq
\phit_{t}=D_{\parallel}\phit_{x x}
+D_{\perp}\nabla^2_{\perp}\phit
+N_{\parallel}(\phit_{x})^2+A\phit_{x x x},\lb{kbperp}
\eeq

\section{Non-isotropic Kuramoto-Sivashinsky Equation}
\quad The diffusion coefficient $D_{\perp}$ in the n-Burgers equation
\fr{nib} is assumed to be small so that
\beq
D_{\perp}\propto \la \hat{U}\cdot DU_{0,\theta} \ra \sim O(\eps^2).\lb{ks1}
\eeq
and  net diffusions along both $x$ and $\rvp$ directions are
nevertheless the same  order of magnitude, that is,
\beq
\partial^2_x/\nabla^2_{\perp} \sim O(\eps^2).\lb{scaleks}
\eeq
 Eq.\fr{ks1} guarantees existence of a periodic vector $W(\theta)$
such that
\beq
LW(\theta)=DU_{0,\theta}+ O(\eps^2).\lb{ks2}
\eeq
 Suppose that $\delta, \tilde U$, and $U$ are expanded as
\beqa
\delta&=&\eps^3\{P_1(x,\rvp,t)+\eps P_2(x,\rvp,t)+\cdots\},\lb{ksdlt}\\
\tilde U&=&\eps^4\{\tilde U_2(\theta)+\eps \tilde U_3(\theta,x,\rvp,t)
+\cdots\},\lb{kstu}\\
U&=&U_0(k,\theta)+\eps^3( U_1+\eps U_2+\eps^2 U_3+\cdots), \lb{ksu}
\eeqa
 then perturbed equations up to $O(\eps^5)$ give the same equations as
 \fr{kb31} and \fr{kb32} for $U_j \quad (j=1,2,3)$.
\beqa
\tilde{L}U_j&=&0,\lb{kshomo}\\
U_j&=&P_j(x,\rvp,t)U_{0,\theta}(\theta)+P_{j,x}(x,\rvp,t)U_{0,k}(\theta)
+\tilde U_j,
\lb{kssol0}
\eeqa
Noting $\tU_1=0$ and following  \fr{nb8}, \fr{nb10}, and \fr{burp}
, we have \fr{bg1} with $P_{1,x}=0$ due to the assumption \fr{scaleks}
,that is,
\beq
U_1=P_1(\rvp)U_{0,\theta},\quad  P_1=\nabla_{\perp}P_1\cdot \rvp.
\lb{kssol1}
\eeq
For $j=2$, substituting \fr{kssol0}
into \fr{kshomo}, we have
\beqa
L \tilde U_2(\theta) &=&\nabla_{\perp}^2P_2DU_{0,\theta}
 +P_{2,xx}(MU_{0,k}+DU_{0,\theta})
+\nabla^2P_{2,x}DU_{0,k}\non\\&&
-P_{2,t}U_{0,\theta}
-P_{2,xt}U_{0,k},\lb{kskou4}
\eeqa
Since $\tilde U_2$ is a function of $\theta$ only,
 all the coefficients of functions of $\theta$ in the RHS of \fr{kskou4}
must be constant.
The consistent renormalization condition and
\fr{kssol1} implies that
a degee of polynomial $P_2$ should be two with respect to $\rvp$
and one with respect to $x$ and $t$ at most,i.e.
$$P_{2,xx}=\nabla^2P_{2,x}=P_{2,xt}=0.$$
The first term of the RHS of \fr{kskou4} does not cause a secular
behaviour of $\tilde U_2$
owing to \fr{ks2} and so $P_{2,t}=0$ is necessary for a perodic
solution $\tilde U_2$.
Thus we have
\beqa
\tilde U_2(\theta)&=&S_2W(\theta),\quad S_2=
\nabla^2_{\perp} P_2=\mbox{constant},
\lb{kss2}\\
P_2&=&P_{2,x}x+(\rvp \rvp: \nabla_{\perp} \nabla_{\perp})P_2/2. \lb{ksp2}
\eeqa
where all the coefficients of monomials in \fr{ksp2} are arbitrary
constants.\\
For $j=3$, we  have the same equation as \fr{kskou4},where the suffix 2
is replaced by 3.
\beqa
L \tilde U_3(\theta) &=&\nabla_{\perp}^2P_3DU_{0,\theta}
 +P_{3,xx}(MU_{0,k}+DU_{0,\theta})
+\nabla^2P_{3,x}DU_{0,k}\non\\&&
-P_{3,t}U_{0,\theta}-P_{3,xt}U_{0,k},\lb{ks10}
\eeqa
The consistent renormalization condition implies that
a degree of polynomial $P_3$ should be three with respect to $\rvp$
 and so $\nabla^2_{\perp} P_3$ in the first term of the RHS of \fr{ks10} is
secular with respect to $\rvp$, while
\beq
P_{3,xx}=0,\quad \nabla^2 P_{3,x}=0,\quad P_{3,xt}=0.\lb{kspp3}
\eeq
Therefore, $\tilde U_3$ takes the form
\beq
\tilde U_3=S_3(\rvp)W(\theta)+\bar U_3(\theta),\lb{kstu3}
\eeq
where $S(\rvp)$ is a polynomial of degree one with respect to $\rvp$.
Then, \fr{ks10} reads
\beq
L \bar U_3(\theta)=(\nabla_{\perp}^2P_3-S_3)DU_{0,\theta}
-P_{3,t}U_{0,\theta}.\non
\eeq
Here $P_{3,t}=0$ is again necessary for a periodic solution $ U_3(\theta)$
and $\nabla_{\perp}^2P_3-S_3=0$
so that a secular term $S_3$ is automatically elliminated
as soon as the phase is renormalized as shown in the last paragraph
of this section. Now, we set
$\bar U_3(\theta)=0$ without loss of generality and have
\beqa
S_3&=&\nabla_{\perp}^2 P_3=\nabla_{\perp}S_3\cdot \rvp,\lb{kss3}\\
P_3&=&x(\rvp:\nabla_{\perp})P_{3,x}+
(\rvp \rvp \rvp:\nabla_{\perp}\nabla_{\perp}\nabla_{\perp})P_3/3!,\lb{ksp3}
\eeqa
where $\nabla_{\perp}S_3, \nabla_{\perp} P_{3,x}$ and $\nabla_{\perp}
\nabla_{\perp}\nabla_{\perp}P_3$ are arbitrary constants.\\
A nonlinear term enters in the perturbed equation to $O(\eps^6)$:
\beqa
\tilde{L}U_4&=&(1/2)F'':U^2_1,\lb{ks15}\\
U_4&=&P_4(x,\rvp,t)U_{0,\theta}+P_{4,x}(x,\rvp,t)U_{0,k}
+(1/2)P^2_1(\rvp)U_{0,\theta \theta}\non\\
&+&\tilde U_4(\theta,x,\rvp,t),\lb{ks16}
\eeqa
from which, we get
\beqa
\tilde L \tilde{U}_4&=&\nabla^2P_4DU_{0,\theta}
+P_{4,xx}MU_{0,k}+\nabla^2P_{4,x}DU_{0,k}\non\\
&&+(1/2)\nabla^2P^2_1DU_{0,\theta \theta}-P_{4,t}U_{0,\theta}
-P_{4,xt}U_{0,k}.\lb{ks17}
\eeqa
The similar discussion leading to \fr{kstu3} yields
\beq
\tU_4(\theta,x,\rvp,t)=S_4(x,\rvp)W(\theta)+ \bar U_4(\theta),\lb{kstu4}
\eeq
where
\beq
S_4=S_{4,x}x+(\rvp \rvp: \nabla_{\perp} \nabla_{\perp})S_4/2,\non
\eeq
and $S_{4,x}$ and  $\nabla_{\perp} \nabla_{\perp}S_4$ are arbitrary
constants.
Then \fr{ks17} is rewritten as
\beqa
L\bar{U}_4(\theta)&=&-(S_4-\nabla^2P_4)DU_{0,\theta}
+P_{4,xx}MU_{0,k}
+S_{4,x}MW\non\\
&&+\nabla^2P_{4,x}DU_{0,k}+(1/2)\nabla^2P^2_1DU_{0,\theta \theta}
+\nabla^2S_4DW\non\\
&&-P_{4,t}U_{0,\theta}-P_{4,xt}U_{0,k},\non
\eeqa
which is an equation for $\bar{U}_4(\theta)$ and
all the coefficients of functions of $\theta$ should be constant.
\beq
S_4-\nabla^2P_4=c_1,\quad P_{4,xx}=c_2,\quad \nabla^2P_{4,x}=c_2,
\quad P_{4,t}=c_4, \lb{ksp4}
\eeq
and $c_5=P_{4,xt}=0$ holds again.
The compatibility condition for $\bar U_4(\theta)$ requires
\beqa
&&(\nabla^2P_2/\eps^2)\la \hat{U}\cdot DU_{0,\theta}\ra
+P_{4,x x}\la \hat{U}\cdot MU_{0,k}\ra
+\nabla^2_{\perp}P_{4,x}\la \hat{U}\cdot MW(\theta)\ra \non\\
&&+\nabla^2_{\perp}P_{4,x}\la \hat{U}\cdot DU_{0,k}\ra
+|\nabla_{\perp}P_1|^2\la \hat{U}\cdot DU_{0,\theta \theta}\ra
+\nabla^4_{\perp}P_4\la \hat{U}\cdot DW(\theta)\ra\non\\
&&-P_{4,t}\la \hat{U}\cdot U_{0,\theta}\ra =0.\lb{comks}
\eeqa
 A renormaliaztion transformation to $O(\eps^6)$ is
\beq
\phit(x,\rvp,t)=\phi+\eps^3\{P_1+\eps P_2+
\eps^2 P_3+\eps^3P_4\},\lb{ksnorm}
\eeq
which gives, noting \fr{kssol1},\fr{ksp2},\fr{ksp3}, and \fr{ksp4},
\beqa
\nabla_{\perp} \phit&=&\eps^3\nabla_{\perp}P_1,\quad
\phit_{x x}=\eps^6P_{4,xx},\quad
\nabla^2_{\perp}\phit
=\eps^4\nabla^2_{\perp}P_2,\non\\
\nabla^2_{\perp}\phit_{x}
&=&\eps^6 \nabla^2_{\perp}P_{4,x},\quad
\nabla_{\perp}^4 \phit=\eps^6 \nabla_{\perp}^4P_4,\quad
\phit_{t}=\eps^6P_{4,t}.\lb{rgks}
\eeqa
Substituting \fr{rgks} into \fr{comks},  we arrive at
 a non-isotropic Kuramoto-Sivashinsky (n-K-S) equation:
\beq
\phit_{t}=D'_{\parallel}\phit_{x x}+D_{\perp}\nabla^2_{\perp}\phit
+N_{\perp}|\nabla_{\perp}\phit|^2+E\nabla^4_{\perp}\phit
+G\nabla^2_{\perp}\phit_{x},\lb{ks24}
\eeq
where
\beqa
E&=&\la \hat{U}\cdot DW(\theta)\ra
 /\la \hat{U}\cdot U_{0,\theta}\ra,\non\\
G
&=&A'+H \non\\
A'&=&\la \hat{U}\cdot DU_{0,k}\ra /\la \hat{U}\cdot U_{0,\theta}\ra,
\quad H=\la \hat{U}\cdot MW(\theta)\ra
/\la \hat{U}\cdot U_{0,\theta}\ra .\lb{ks25}
\eeqa
\\
Secular terms in $\tU$ are also eliminated by the present
renormalization procedure.
To $O(\eps^6)$, \fr{kss2}, \fr{kstu3}, and \fr{kstu4} give
\beq
\tU=\eps^4\{S_2+\eps S_3(\rvp)+\eps^2S_4(x,\rvp)\}W(\theta)+
\eps^6 \bar U_4(\theta),\lb{kbr}
\eeq
where  secular terms  $S_3$ and $S_4$ are removed by introducing
a renormalized $S_2$ such that
\beq
\tilde S_2(x,\rvp,t)=S_2+\eps S_3(\rvp)+\eps^2S_4(x,\rvp).\lb{rren}
\eeq
This renormalization transformation is found to be consistent
with that to the phase $\phi$ \fr{ksnorm}, since \fr{kss2}, \fr{kss3},
\fr{rgks}, and \fr{ksp4} with $c_1=0$ imply
$\tilde S_2=\nabla^2_{\perp}\phit/\eps^4$.

\section{ Benney Equation in Three Dimension}
\quad In addition to the assumption \fr{ks1},
the wave number $k$ is also assumed to be as small as $O(\eps)$.
Furthermore it may be reasonable to assume that
$\omega(k)$ and $U_{0}(k,\theta)$ are functions of $k^2$,
which is satisfied in the case of the complex Ginzburg-Landau equation
analyzed in the next section.
Then the following estimates hold
\beq
\dot \omega \sim O(\eps),\quad U_{0,k} \sim O(\eps),
\eeq
and
\beqa
&&A'\propto \la \hat{U}\cdot DU_{0,k}\ra  \sim O(\eps)
,\quad  B \propto \la \hat{U}\cdot MV(\theta)\ra \sim O(\eps) ,\\
&& D'_{\parallel} \propto \la \hat{U}\cdot MU_{0,k}\ra
\sim O(\eps^2),
\eeqa
or
\beq
A=A'+B \sim O(\eps)
,\quad D_{\parallel}=D_{\perp}+D'_{\parallel}\sim O(\eps^2),
\eeq
namely both coefficients of diffusion $D_{\perp}$ and $D_{\parallel}$
are as small as $O(\eps^2)$ while the coefficient of dispersion
$A$ is $O(\eps)$.
Here, the linearized operator $L$ and $\tilde L$ are rearranged as
\beqa
&&L=L'-k^2D\partial^2_\theta,\\
&&L'=-\omega \partial_\theta-F'(U_0)\cdot,
\eeqa
and
\beqa
&&\tilde L =\tilde L' -M\partial_x-k^2D\partial^2_\theta,\\
&&\tilde L' =\partial_t + L'+\nabla^2D.
\eeqa
Then \fr{nb3}--\fr{nb7} become
\beqa
L'U_{0,\theta}&=&k^2D\partial^2_\theta U_{0,\theta},
\lb{rdif3}\\
L'U_{0,\theta \theta}&=&F'': U^2_{0,\theta}
+k^2D\partial^2_\theta U_{0,\theta \theta},\lb{rdif4}\\
L'U_{0,k}&=& MU_{0,\theta}+k^2D\partial^2_\theta U_{0,k},\lb{ph13}\\
L'U_{0,kk}&=&F'': U^2_{0,k}+2MU_{0,\theta k}+2DU_{0,\theta \theta}\non\\
&+&\ddot{\omega}U_{0,\theta}+k^2D\partial^2_\theta U_{0,kk},\lb{ph14}\\
L'U_{0,\theta k}&=&MU_{0,\theta \theta}+F'': U_{0,\theta}U_{0,k}
+k^2D\partial^2_\theta U_{0,\theta k}.\lb{ph15}
\eeqa
Suppose that $\delta,\tilde U$, and $U$ are expanded in the same forms
as \fr{ksdlt}, \fr{kstu}, and \fr{ksu}, then
 up to $O(\eps^5)$ perturbed secular fields $U_j \quad (j=1,2,3)$  obey
\beqa
\tilde{L}'U_j&=&M\partial_x U_{j-1}, \lb{bn1}\\
U_j&=&P_j(x,\rvp,t)U_{0,\theta}+P_{j-1,x}U_{0,k}
+\tU_j ,\lb{bn2}
\eeqa
where $P_0=\tU_1=0$.\\
For $j=1$, \fr{bn1} and \fr{bn2} give
\beq
P_{1,t}U_{0,\theta}-\nabla^2P_1DU_{0,\theta}=0,
\eeq
which implies
\beq
P_{1,t}=0,\quad \nabla^2P_1=0,
\eeq
and
\beq
P_1=P_{1,x}x+\nabla_{\perp}P_1\cdot \rvp. \lb{bnp1}
\eeq
To $j=2$, \fr{bn1} and \fr{bn2} give
\beq
\tilde L' \tilde U_2=-P_{2,t}U_{0,\theta}+\nabla^2P_2DU_{0,\theta}.
\eeq
The similar discussion as in the previous sections yields
\beq
\tilde U_2=S_2W(\theta), \quad S_2=\nabla^2P_2=\mbox{constant}
,\quad P_{2,t}=0,
\eeq
and
\beq
P_2=P_{2,xx}x^2/2+x( \rvp:\nabla_{\perp})P_{2,x}+(\rvp \rvp:\nabla_{\perp}
\nabla_{\perp})P_2/2, \lb{bnp2}
\eeq
where $P_{2,xx},\nabla_{\perp}P_{2,x}$ and $(\nabla_{\perp})^2P_2$ are
arbitrary constants.\\
Similarly, to $j=3$, we have
\beq
\tilde U_3=S_3W(\theta),\quad S_3=\nabla^2P_3=S_{3,x}x+\nabla_{\perp}S_3
 \cdot \rvp,\quad P_{3,t}=0,
\eeq
and
\beqa
P_3&=&P_{3,xxx}x^3/3!+x^2(\rvp:\nabla_{\perp})P_{3,xx}/2
+x(\rvp \rvp:\nabla_{\perp}\nabla_{\perp})P_{3,x}/2\non\\
&&+(\rvp \rvp \rvp:\nabla_{\perp}\nabla_{\perp}\nabla_{\perp})P_3/3!,
\lb{bnp3}
\eeqa
which is a polynomial of degree three and all the coefficients
are arbitrary constants\\
\quad To $O(\eps^6)$, a nonlinear term appears as
\beqa
&&\tilde L' U_4=(1/2)F'':U^2_1+M\partial_xU_3,\lb{bn6}\\
&&U_4=P_4U_{0,\theta}+P_{3,x}U_{0,k}+(1/2)P^2_1U_{0,\theta \theta}
+\tilde U_4(\theta,x,\rvp,t).\lb{bn66}
\eeqa
Substituting \fr{bn66} into \fr{bn6}, we get
\beqa
\tilde L' \tilde U_4&=&\nabla^2P_4DU_{0,\theta}+\nabla^2P_{3,x}DU_{0,k}
+P_{2,xx}MU_{0,k}\non\\
&&+\nabla^2(P^2_1/2)DU_{0,\theta \theta} -P_{4,t}U_{0,\theta}.
\eeqa
Let set
\beq
\tilde U_4(\theta)=S_4W(\theta)+\bar U_4 (\theta),
\eeq
where
\beq
S_4=S_{4,xx}x^2/2+x(\rvp:\nabla_{\perp})S_{4,x}+(\rvp \rvp:\nabla_{\perp} \nabla_{\perp})S_4/2,
\eeq
which is a polynomial of degee two,
then we have
\beqa
L'\bar U_4&=&-(S_4-\nabla^2P_4)DU_{0,\theta}+\nabla^2P_{3,x}DU_{0,k}
+P_{2,xx}MU_{0,k}\non\\
&&-\nabla^2S_4DW-S_{3,x}MW+\nabla^2(P^2_1/2)DU_{0,\theta}
-P_{4,t}U_{0,\theta}. \lb{bnu4}
\eeqa
This equation is valid only when
all the coefficients of functions of $\theta$ in \fr{bnu4} are constant
and the compatibility condition gives
\beqa
&&(\nabla^2P_2/\eps^2)\la \hat{U}\cdot DU_{0,\theta}\ra
+P_{2,x x}\la \hat{U}\cdot MU_{0,k}\ra +\nabla^2P_{3,x}\la \hat{U}\cdot
 MW\ra\non\\
&&+\nabla^2P_{3,x}\la \hat{U}\cdot DU_{0,k}\ra
+|\nabla P_1|^2\la \hat{U}\cdot DU_{0,\theta \theta}\ra
+\nabla^4 P_4\la \hat{U}\cdot DW\ra\non\\&&-P_{4,t}\la \hat{U}\cdot
U_{0,\theta}\ra =0.\lb{combn}
\eeqa
Now a renormalization transformation to $O(\eps^6)$ takes
the same form as \fr{ksnorm}.
From \fr{ksnorm},\fr{bnp1},\fr{bnp2},\fr{bnp3} and \fr{bnu4} we obtain
\beqa
&&\phit_{x}=\eps^3P_{1,x},\quad
\nabla_{\perp} \phit=\eps^3\nabla_{\perp}P_1,\quad
\phit_{x x}=\eps^4P_{2,xx},\quad
\nabla^2_{\perp}\phit=\eps^4\nabla^2_{\perp}P_2, \non\\
&&\phit_{xxx}=\eps^5P_{3,xxx},\quad
\nabla^2_{\perp}\phit_{x}=\eps^5\nabla^2_{\perp}P_{3,x},\quad
\phit_{xxxx}=\eps^6P_{4,xxxx},\quad \nabla^4_{\perp}\phit=\eps^6 \nabla^4_{\perp}P_4,\non\\
&&\phit_{t}=\eps^6P_{4,t}.\lb{rgbn}
\eeqa
Substituting \fr{rgbn} into \fr{combn},
we arrive at a generalized Benney equation in the three dimensional space:
\beq
\phit_{t}=D_{\parallel}\phit_{xx}+D_{\perp}\nabla^2_{\perp}\phit
+G(\phit_{xxx}+\nabla^2_{\perp}\phit_{x})+E\nabla^4 \phit+
N_{\perp}|\nabla \phit|^2.
\eeq
As in the case of the n-K-S equation, secular terms in $\tU$ are
also automatically removed when $\phi$ is renormalized.

\section{Complex Ginzburg-Landau Equation}
\quad As an application of the previous results, let us calculate
explicitly various coefficients of the
 phase equations for
the complex Ginzburg-Landau (cGL) equation.
\beq
\left\{ \begin{array}{l} \mit\Psi_{,t}=\gamma{\mit\Psi}-\beta|{\mit\Psi}|^2
{\mit\Psi}
+\alpha\nabla^2{\mit\Psi}\\
\bar{\mit\Psi}_{,t}=\gamma\bar{\mit\Psi}-\bar{\beta}|{\mit\Psi}|^2{\bar{\mit
\Psi}}
+{\bar\alpha}\nabla^2{\bar{\mit\Psi}}, \end{array} \right.   \lb{gl2}
\eeq
where $\Psi$ is a complex variable, $\gamma$ is a real constant,
$\alpha$ and $\beta$ are complex constants. The bar denotes complex
conjugation.
Setting
\beq
U=
\left( \begin{array}{c} \mit\Psi\\ \bar{\mit\Psi} \end{array} \right),\quad
F=
\left( \begin{array}{c} \gamma\mit\Psi-\beta|\mit\Psi|^2\mit\Psi\\
\gamma\bar{\mit\Psi}-\bar{\beta}|{\mit\Psi}|^2 \bar{\mit\Psi}
 \end{array} \right),\quad
D=
\left( \begin{array}{cc} \alpha & 0 \\ 0 & \bar\alpha \end{array} \right),
\lb{gl1}
\eeq
the cGL equation is transformed into the standard form of
a reaction-diffusion system \fr{nb1}.
As a periodically oscillating solution of \fr{gl2}, we take
\beq
U_0=
\left( \begin{array}{c} \mit\Psi_0 \\ \bar{\mit\Psi}_0 \end{array} \right)
=a(k) \left( \begin{array}{c} e^{ i\theta} \\ e^{-i\theta} \end{array}
\right)
,\lb{gl3}
\eeq
where a real amptitude $a(k)$ and a frequency $\omega(k)$ satisfy
the following dispersion relation
\beq
-i\omega=\gamma-\beta a^2-\alpha k^2 ,\lb{gl4}
\eeq
which  $\omega(k)$ and $a(k)$ are functions of $k^2$ as speculated in
section 6.
\quad Since the ajoint operator $L^\dagger$ of
$L$ defined in \fr{nb3} is
\beq
L^\dagger=\omega \partial_\theta-F'^\dagger(U_0)
-D^\dagger k^2 \partial^2_\theta,\lb{gl5}
\eeq
where
\beq
F'^\dagger (U_0)= \left(
\begin{array}{cc} \gamma-2\bar{\beta}a^2 & -\beta {\mit{\Psi}}^2_0 \\
-\bar{\beta}{\bar{\mit\Psi}}^2_0 & \gamma-2\beta a^2 \end{array} \right),\quad
D^\dagger= \left( \begin{array}{cc}
\bar{\alpha} & 0 \\
0 & \alpha \end{array} \right),\lb{gl6}
\eeq
we explicitly find the null vector of $L^\dagger$ as
\beq
\hat{U}=i/2\beta'a \left( \begin{array}{c}
\beta e^{i\theta} \\
-\bar{\beta} e^{-i\theta} \end{array} \right),\lb{gl7}
\eeq
with the normalization $\la \hat{U} \cdot U_{0,\theta} \ra =1 $ and
 $\beta=\beta'+i\beta''$.
Various differentials of $U_0$ are also explicitly given by
\beqa
&&U_{0,\theta}=a \left( \begin{array}{c} i e^{i\theta}\\ -i e^{-i\theta}
\end{array} \right),\quad
U_{0,k}=\dot{a} \left( \begin{array}{c} e^{i\theta}\\ e^{-i\theta}
\end{array} \right),\quad
U_{0,\theta \theta}=-a \left( \begin{array}{c} e^{i\theta}\\ e^{-i\theta}
\end{array} \right), \non\\
&&U_{0,\theta k}=\dot{a} \left( \begin{array}{c} ie^{i\theta}\\
-ie^{-i\theta} \end{array} \right),\quad
U_{0,k k}= \ddot{a} \left( \begin{array}{c} e^{i\theta}\\ e^{-i\theta}
\end{array} \right).\lb{gl8}
\eeqa
Using \fr{gl7} and \fr{gl8}, the coefficients of diffusion and nonlinearity
are calculated as
\beqa
D_{\perp}&=&\la \hat{U} \cdot DU_{0,\theta} \ra= Re(\alpha \bar \beta)/\beta',
\lb{gl12}\\
D'_{\parallel}&=&\la \hat{U} \cdot MU_{0,k} \ra
 =(-\dot{a}/\beta'a)\{\dot{\omega}\beta''-2kRe(\alpha \bar{\beta})\}\non\\
&=&-2\dot{a}^2|\beta|^2/\beta',\lb{gl13}\\
N_{\perp}&=&\la \hat{U} \cdot DU_{0,\theta \theta} \ra
=-Im(\alpha \bar{\beta})/\beta'=-\ddot{\omega}/2=N_{\parallel},\lb{gl14}
\eeqa
where $N_{\parallel}$ happens to be identical with $N_{\perp}$ in this case.
The coefficient of dispersion in the K-dV-Burgers equation is
calculated as follows.
The assumption \fr{kb1} reads
\beq
D_{\parallel}=\{Re(\alpha \bar{\beta})-2\dot{a}^2|\beta|^2\}/\beta'
\sim O(\eps),\lb{gl15}
\eeq
and
\beq
Re(\alpha \bar{\beta})\sim2\dot{a}^2|\beta|^2.\lb{gl16}
\eeq
Then $V$ in \fr{kb2} is obtained as
\beq
V(\theta)=(Im(\bar{\alpha} \beta)/2a\dot{a}|\beta|^2)U_{0,k},\lb{gl18}
\eeq
and we have the coefficient of dispersion
\beqa
A'&=&\la \hat{U}\cdot DU_{0,k} \ra
=\dot{a}Im(\alpha \bar{\beta})/a\beta',\lb{gl19}\\
B&=&\la \hat{U}\cdot MV(\theta)\ra
=(Im(\bar{\alpha} \beta)/2a\dot{a}|\beta|^2) \la \hat{U}\cdot MU_{0,k} \ra
\non\\&=&-\dot{a}Im(\bar{\alpha}\beta)/a\beta',\lb{gl20}\\
A&=&B+A'=2\dot{a}Im(\alpha \bar{\beta})/a\beta'.\lb{gl21}
\eeqa

From the Kuramoto-Sivashinsky scale \fr{ks1}, we have
\beq
D_{\perp}= \la \hat{U}\cdot DU_{0,\theta} \ra =Re(\alpha \bar{\beta})/\beta'
\sim Re(\alpha \bar{\beta})\sim O(\eps^2).\lb{gl22}
\eeq
Then $W$ in \fr{ks2} is found to have the same expression as $V$ since
the forcing terms in Eqs.\fr{kb2} and \fr{ks2} have the same 
non-seculer component under the conditions \fr{gl15} and \fr{gl22}.

The coefficients of the K-S equation are obtained as
\beqa
E&=&\la \hat{U} \cdot DW(\theta)\ra
=(Im(\bar{\alpha} \beta)/2a\dot{a}|\beta|^2)\la \hat{U}\cdot DU_{0,k}\ra\non\\
&=&-(Im(\bar{\alpha}\beta))^2/2a^2\beta'|\beta|^2,\lb{gl26}\\
H&=&\la \hat{U}\cdot MW(\theta)\ra =B,\lb{gl27}\\
G&=&A'+H=A'+B=A.\lb{gl28}
\eeqa
All the coefficients of the Benney equation are given in terms of
those of the K-B equation and the K-S equation.
The isotropic part of coefficients of the n-Burgers equation and
the K-S equation agrees with the result in \cite{maru}.
\section{Concluding Remarks}
Let us compare the present derivation of phase equations with
a possible derivation by means of the reductive perturbation (RP)
method \cite{tani}
 or the multi space-time scale method, although the latter
derivation has not been accomplished yet.
The initial setting of perturbation \fr{solution} (and \fr{kappa}) and
auxiliary conditions
\fr{kb1} and \fr{ks1} are same for both derivations.
In addition to the initial setting and the auxiliary condition,
the RG method assumes
a naive expansion of secular solutions such as \fr{delex1}--\fr{nbexu}
 and \fr{kb}--\fr{kbux} etc. . Then straightforward calculations of
 secular terms with the aid of the consistent renormalization condition
 and the RG procedure of the Lie approach lead to the final results.
 The type of derived phase equations depends only on a specific form
 of expansion of secular terms and the auxiliary condition.
 On the other hand, the RP method would requires specific
 scalings for not only perturbed fields but also independent variables,
 which are available after a derived equation is set up.
 It should be mentioned that the present RG method relies on an
 explicit secular solution although the RP method would not.
 However, the step to obtain an explicit secular solution may
 be largely skipped as far as the final results are concerned or
 possibly by the proto-RG approach developed recently \cite{nos}.

\end{document}